\documentclass[11pt]{article}

\usepackage[margin=1in]{geometry}
\usepackage{setspace}
\usepackage{amsmath,amssymb,amsthm}
\usepackage{bm}
\usepackage{graphicx}
\usepackage{color}
\usepackage{hyperref}

\usepackage{algorithm}
\usepackage{algpseudocode}

\newcommand{\R}{\mathbb{R}}

\newcommand{\Var}{\mathrm{Var}}

\newcommand{\veps}{\varepsilon}

\newcommand{\dBij}{B_{i,j}}
\newcommand{\dBik}{B_{i,k}}
\newcommand{\dAij}{A_{i,j}}
\newcommand{\dAik}{A_{i,k}}
\newcommand{\tSigma}{\widetilde{\Sigma}}
\newcommand{\logdet}{\mathtt{logdet}}
\newcommand{\dlogdetj}{\mathtt{dlogdetj}}
\newcommand{\XSX}{\mathtt{XSX}}
\newcommand{\XSy}{\mathtt{XSy}}
\newcommand{\ySy}{\mathtt{ySy}}
\newcommand{\dXSXj}{\mathtt{dXSXj}}
\newcommand{\dXSyj}{\mathtt{dXSyj}}
\newcommand{\dySyj}{\mathtt{dySyj}}
\newcommand{\Trjk}{\mathtt{Trjk}}
\newcommand{\betatheta}{\widehat{\beta}(\theta)}
\newcommand{\Tr}{\mbox{Tr}}

\usepackage[round]{natbib}

\begin{document}

\begin{center}{\LARGE Gaussian Process Learning via Fisher Scoring

\vspace{6pt}

of Vecchia's Approximation }

\vspace{16pt}

{Joseph Guinness}
\vspace{8pt}

\textit{Cornell University, Department of Statistics and Data Science}
\vspace{16pt}

\textbf{Abstract}
\end{center}

\noindent We derive a single pass algorithm for computing the gradient and
Fisher information of Vecchia's Gaussian process loglikelihood approximation, which
provides a computationally efficient means for applying the
Fisher scoring algorithm for maximizing the loglikelihood. The
advantages of the optimization techniques are demonstrated in numerical examples
and in an application to Argo ocean temperature data. The new methods are more accurate and much faster than an optimization
method that uses only function evaluations, especially when the covariance
function has many parameters. This allows practitioners to fit nonstationary
models to large spatial and spatial-temporal datasets.

\section{Introduction}\label{introduction}

The Gaussian process model is an indispensible tool for the analysis of spatial and spatial-temporal datasets and has become increasingly popular as a general-purpose model for functions. Because of its high computational burden, researchers have devoted substantial effort to developing numerical approximations for Gaussian process computations. Much of the work focuses on efficient approximation of the likelihood function. Fast likelihood evaluations are crucial for optimization procedures that require many evaluations of the likelihood, such as the default Nelder-Mead algorithm \citep{nelder1965simplex} in the R \verb!optim! function. The likelihood must be repeatedly evaluated in MCMC algorithms as well.

Compared to the amount of literature on efficient likelihood approximations, there has been considerably less development of techniques for numerically maximizing the likelihood (see \cite{geoga2018scalable} for one recent example). This article aims to address the disparity by providing:
\begin{enumerate}
\item Formulas for evaluating the gradient and Fisher information for Vecchia's likelihood approximation in a single pass through the data, so that the Fisher scoring algorithm can be applied. Fisher scoring is a modification of the Newton-Raphson optimization method, replacing the Hessian matrix with the Fisher information matrix.
\item Numerical examples with simulated and real data demonstrating the practical advantages that the new techniques provide over an optimizer that uses function evaluations alone.
\end{enumerate}

Among the sea of Gaussian process approximations proposed over the past several decades, Vecchia's approximation \citep{vecchia1988estimation} has emerged as a leader. It can be computed in linear time and with linear memory burden, and it can be parallelized. Maximizing the approximation corresponds to solving a set of unbiased estimating equations, leading to desirable statistical properties \citep{stein2004approximating}. It is general in that it does not require gridded data nor a stationary model assumption. The approximation forms a valid multivariate normal model, and so it can be used for simulation and conditional simulation. As an approximation to the target model, it is highly accurate relative to competitors \citep{guinness2018permutation}. Vecchia's approximation also forms a conceptual hub in the space of Gaussian process approximations, since a generalization includes many well-known approximations as special cases \citep{katzfuss2017general}. Lastly, there are publicly available R packages implementing it \citep{spNNGP,GpGp}.

The numerical examples in this paper show that, in realistic data and model scenarios, the new techniques offer significant computational advantages over default optimization techniques. Although it is more expensive to evaluate the gradient and Fisher information in addition to the likelihood, the Fisher scoring algorithm converges in a small number of iterations, leading to a large advantage in total computing time over an optimization method that uses only the likelihood. For isotropic Mat\'ern models, the speedup is roughly 2 to 4 times, and on more complicated models with more parameters, the new techniques can be more than 40 times faster. This is a significant practical improvement that will be attractive to practitioners choosing among various methods.

\section{Background}

Let $s_1,\ldots,s_n$ be locations in a domain $D$. At each $s_i$, we observe a scalar response $y_i$, collected into column vector $y = (y_1,\ldots,y_n)^T$. Along with the response, we observe covariates $x_i = (x_{i1},\ldots,x_{ip})$ collected into an $n\times p$ design matrix $X$. In the Gaussian process, we model $y$ as a multivariate normal vector $Y$ with expected value $E(Y) = X \beta$ ($\beta \in \R^p$), and covariance matrix $E( (Y - X\beta)(Y-X\beta)^T ) = \Sigma_\theta$, where the $(i,j)$ entry of $\Sigma_\theta$ is
$K_\theta( s_i, s_j)$. The function
$K_\theta$ is positive definite on $D \times D$ and depends on covariance parameters $\theta$. The loglikelihood for $\beta$ and $\theta$ is
\begin{align}
\log f_{\beta,\theta}(y) = -\frac{n}{2} \log (2\pi) - \frac{1}{2} \log \det \Sigma_\theta - \frac{1}{2}( y - X \beta)^T \Sigma_\theta^{-1} (y - X\beta).
\end{align}
Unless $\Sigma_\theta$ has some exploitable structure, evaluation of the loglikelihood involves storing the $n^2$ entries of $\Sigma_\theta$ and performing $O(n^3)$ floating point operations to obtain the Cholesky factor of $\Sigma_\theta$, both of which are computationally prohibitive when $n$ is large.

Vecchia's loglikelihood approximation is a modification of the conditional representation of a joint density function. Let $g(1) = \emptyset$, $g(i) \subset (1,\ldots,i-1)$ and $y_{g(i)}$ be the corresponding subvector of $y$. Vecchia's loglikelihood approximation is
\begin{align}
{\ell}(\beta,\theta) = \sum_{i=1}^n \log f_{\beta,\theta}(y_i | y_{g(i)} ),
\end{align}
leading to computational savings when $|g(i)|$ is small. As mentioned in the introduction, Vecchia's likelihood approximation corresponds to a valid multivariate normal distribution with mean $X\beta$ and a covariance matrix $\tSigma_\theta$. To motivate why obtaining the gradient and Fisher information poses an analytical challenge, consider the partial derivative of Vecchia's loglikelihood with respect to $\theta_j$:
\begin{align}
\frac{\partial \ell(\beta,\theta)}{\partial \theta_j} = \frac{1}{2}(y - X\beta)^T \tSigma_\theta^{-1} \frac{\partial \tSigma_\theta}{\partial \theta_j} \tSigma_\theta^{-1}(y - X\beta) - \frac{1}{2} \mbox{Tr}\left( \tSigma_\theta^{-1} \frac{\partial \tSigma_\theta}{\partial \theta_j} \right),
\end{align}
where $(\partial \tSigma_\theta/ \partial \theta_j)$ is an $n \times n$ matrix of partial derivatives of $\tSigma_\theta$ with respect to $\theta_j$. Not only is $\partial\tSigma_\theta/\partial \theta_j$ too large to store in memory, the covariances $\tSigma_\theta$ are not easily computable, nor are their partial derivatives. In the next section, we outline a simple reframing of Vecchia's likelihood that leads to a computationally tractable method of evaluating the gradient and Fisher information.

\section{Derivations for Single Pass Algorithm}

To derive formulas for the gradient and Fisher information, it is helpful to rewrite the conditional likelihoods in terms of marginals. To this end, define $u_i = y_{g(i)}$ and $v_i = (y_{g(i)}, y_i)$. Define the design matrices for $u_i$ and $v_i$, respectively, as $Q_i$ and $R_i$, and define the covariance matrices for $u_i$ and $v_i$, respectively as $A_i$ and $B_i$ (suppressing dependence on $\theta$). The notation is chosen to follow the mnemonic device that the first of the two letters alphabetically is a subvector or submatrix of the second letter. Vecchia's loglikelihood can then be rewritten as
\begin{align}
{\ell}(\beta,\theta) =& \sum_{i=1}^m \log f_{\beta,\theta}(v_i) - \log f_{\beta,\theta}(u_i) \\
=&  -\frac{1}{2} \sum_{i=1}^n \left[ \log \det B_i  - \log \det A_i \right] \\
&- \frac{1}{2}\sum_{i=1}^n \left[ (v_i - R_i \beta)^T B_i^{-1} (v_i - R_i \beta ) - (u_i - Q_i \beta)^T A_i^{-1} (u_i - Q_i \beta) \right] - \frac{n}{2}\log(2\pi).
\end{align}
Our proposed algorithm for obtaining the likelihood, gradient, and Fisher information involves computing the following quantities in a single pass through the data.
\begin{align}
\left( \logdet \right) &= \sum_{i=1}^n \left( \log \det B_i - \log \det A_i \right) \\
\left( \dlogdetj \right) &= \sum_{i=1}^n \left( \Tr(B_i^{-1} \dBij) - \Tr(A_i^{-1} \dAij) \right) \\
\left( \ySy \right) &= \sum_{i=1}^n \left( v_i^T B_i^{-1} v_i - u_i^T A_i^{-1} u_i \right) \\
\left( \XSy \right) &= \sum_{i=1}^n \left( R_i^T B_i^{-1} v_i - Q_i^T A_i^{-1} u_i \right) \\
\left( \XSX \right) &= \sum_{i=1}^n \left( R_i^T B_i^{-1} R_i - Q_i^T A_i^{-1} Q_i \right) \\
\left( \dySyj \right) &= -\sum_{i=1}^n \left( v_i^T B_i^{-1} \dBij B_i^{-1} v_i - u_i^T A_i^{-1} \dAij A_i^{-1} u_i \right) \\
\left( \dXSyj \right) &= -\sum_{i=1}^n \left( R_i^T B_i^{-1} \dBij B_i^{-1} v_i - Q_i^T A_i^{-1} \dAij A_i^{-1} u_i \right) \\
\left( \dXSXj \right) &= -\sum_{i=1}^n \left( R_i^T B_i^{-1} \dBij B_i^{-1} R_i - Q_i^T A_i^{-1} \dAij A_i^{-1} Q_i \right) \\
\left( \Trjk \right) &= \sum_{i=1}^n \left[ \Tr(B_i^{-1}\dBij B_i^{-1} \dBik) - \Tr(A_i^{-1}\dAij A_i^{-1} \dAik) \right],
\end{align}
where $\dAij$ and $\dBij$ are the matrices of partial derivatives of $A_i$ and $B_i$, respectively, with respect to $\theta_j$. The quantities having the form $(\mathtt{d*j})$ are simply the partial derivatives of the corresponding quantity $(\mathtt{*})$ with respect to $\theta_j$. Each of these quantities can be updated at each $i=1,\ldots,n$, and so all can be evaluated in a single pass through the data. We refer to them collectively as our single-pass quantities.

\subsection{Profile Likelihood, Gradient, and Fisher Information}

Given covariance parameter $\theta$, denote the maximum Vecchia likelihood estimate of $\beta$ as $\widehat{\beta}(\theta)$. Since $\betatheta$ has a closed form expression (Section \ref{mean_parameters}), we can maximize the profile likelihood $\ell(\betatheta,\theta)$ over $\theta$ alone.
%\begin{align}
%\ell(\betatheta,\theta) =&  -\frac{n}{2}\log(2\pi)
% -\frac{1}{2} \sum_{i=1}^n \left[ \log \det B_i  - \log \det A_i \right] \nonumber \\
%  &- \frac{1}{2}\sum_{i=1}^n \left[ (v_i - R_i \betatheta)^T B_i^{-1} (v_i - R_i \betatheta ) - (u_i - Q_i \betatheta)^T A_i^{-1} (u_i - Q_i \betatheta) \right].
%\end{align}
The profile likelihood can be written in terms of our single-pass quantities as
\begin{align}
\ell(\betatheta,\theta) = - \frac{n}{2}\log(2\pi) -\frac{1}{2}(\logdet) - \frac{1}{2}\left[ (\ySy) - 2(\XSy)\betatheta + \betatheta^T (\XSX) \betatheta \right] .
\end{align}
Therefore the partial derivatives can also be written in terms of the single pass quantities as
\begin{align}
\frac{ \partial  \ell(\betatheta,\theta)}{\partial \theta_j} = &-\frac{1}{2}(\dlogdetj)
 - \frac{1}{2}\left[ (\dySyj) - 2(\dXSyj)\betatheta + \betatheta^T (\dXSXj) \betatheta \right] \nonumber \\
 &- \frac{1}{2}\left[ - 2(\XSy)\frac{ \partial \betatheta}{\partial \theta_j} + 2\betatheta^T (\XSX) \frac{ \partial \betatheta}{\partial \theta_j} \right],
\end{align}
where $(\partial \widehat{\beta}(\theta) / \partial \theta_j )$ is the column vector of partial derivatives of the $p$ entries of $\widehat{\beta}(\theta)$ with respect to covariance parameter $\theta_j$. The Fisher information is
\begin{align*}
\mathcal{I}(\theta)_{jk} = \frac{1}{2}\sum_{i=1}^n \left[ \Tr(B_i^{-1}\dBij B_i^{-1} \dBik) - \Tr(A_i^{-1}\dAij A_i^{-1} \dAik) \right] = \frac{1}{2}\left( \Trjk \right).
\end{align*}
It remains to be shown that $\betatheta$ and $\partial \betatheta/ \partial \theta_j$ can be computed using our single-pass quantities.

\subsection{Mean Parameters}\label{mean_parameters}

The profile likelihood estimate $\betatheta$ satisfies $\partial \ell(\beta,\theta)/\partial \beta_j = 0$ for every $j=1,\ldots,p$. These partial derivatives are
\begin{align}
\begin{bmatrix}
\partial \ell(\beta,\theta)/\partial \beta_1 \\
\vdots \\
\partial \ell(\beta,\theta)/\partial \beta_p
\end{bmatrix} = \sum_{i=1}^n R_i^T B_i^{-1}(v_i - R_i \beta) - Q_i^T A_i^{-1}(u_i - Q_i \beta),
\end{align}
giving the equation
\begin{align}\label{betascore}
\left[ \sum_{i=1}^n \Big( R_i^T B_i^{-1} R_i - Q_i^T A_i^{-1} Q_i \Big) \right] \widehat{\beta}(\theta) = \left[ \sum_{i=1}^n \Big( R_i^T B_i^{-1} v_i - Q_i^T A_i^{-1} u_i \Big) \right].
\end{align}
Therefore, the profile likelihood estimate of $\beta$ is
\begin{align}
\widehat{\beta}(\theta) = (\XSX)^{-1} (\XSy),
\end{align}
a function of our single pass quantities. Taking partial derivatives with respect to $\theta_j$ yields
\begin{align}
\frac{\partial \betatheta }{\partial \theta_j} = (\XSX)^{-1} (\dXSyj) - (\XSX)^{-1}(\dXSXj)(\XSX)^{-1}(\XSy),
\end{align}
also a function of our single pass quantities.

\section{Numerical Studies}

This section contains timing results, comparing the R \verb!optim! implementation of the Nelder-Mead algorithm to the Fisher scoring algorithm. In Fisher scoring, we reject steps that do not increase the loglikelihood, dividing the step size by 2 iteratively. If we cannot increase the loglikelihood along the Fisher step direction, we attempt to step along the gradient. For both Nelder-Mead and Fisher scoring, we first generate estimates using Vecchia's approximation with $|g(i)| = 10$ nearest neighbors, then refine the estimates using $|g(i)| = 30$. In Nelder-Mead, we evaluate only the likelihood, not the gradient and Fisher information. The Fisher scoring algorithm stops when the dot product between the step and the gradient is less than 1e-4. Default stopping criteria were used for Nelder-Mead algorithm. We simulate all datasets from the same model:
\begin{align}
Y(s) = \mu + Z(s) + \veps(s),
\end{align}
where $\mu=0$, $Z$ is a Gaussian process with exponential covariance function $K(s_1,s_2) = \sigma^2 \exp(-\|s_1 - s_2 \|/\alpha)$, and $\veps(s)$ are i.i.d.\ $N(0,\tau^2)$ with $\tau^2 = 0.2$. We take $(\sigma^2, \alpha) = (2,0.3)$. Data are simulated on an evenly spaced grid of 4900 locations on $[0,1]^2$. In addition to the exponential covariance with unknown variance and range, we estimate parameters in three covariance models that generalize the exponential:
\begin{align}
K(s_1,s_2) &= \frac{\sigma^2}{\Gamma(\nu)2^{\nu-1}} \left( \frac{\|s_1 - s_2 \|}{\alpha} \right)^{\nu} \mathcal{K}_{\nu}\left( \frac{\|s_1 - s_2 \|}{\alpha} \right) \\
K(s_1,s_2) &= \frac{\sigma^2}{\Gamma(\nu)2^{\nu-1}} \left( \|Ls_1 - Ls_2 \| \right)^{\nu} \mathcal{K}_{\nu}\left( \|Ls_1 - Ls_2 \|\right) \\
K(s_1,s_2) &= \exp\left( \sum_{j=1}^J b_j ( \phi_j(s_1) + \phi_j(s_2) ) \right)\frac{\sigma^2}{\Gamma(\nu)2^{\nu}} \left( \frac{\|s_1 - s_2) \|}{\alpha} \right)^{\nu} \mathcal{K}_{\nu}\left( \frac{\|s_1 - s_2 \|}{\alpha} \right).
\end{align}
The first is an isotropic Mat\'ern covariance function. The second is a geometrically anisotropic Mat\'ern covariance, with anisotropy parameterized by the $2 \times 2$ lower triangular matrix $L$. The third is a Mat\'ern covariance with a nonstationary variance function. The nonstationary variances are defined in terms of pre-specified known basis functions $\phi_j$ and unknown parameters $b_j$. For identifiability purposes, the $J=8$ basis functions are an orthogonal basis that is also orthogonal to a constant function. The orthogonal basis is formed by applying Gram-Schmidt orthogonalization to a set of Gaussian basis functions.

Excluding $\mu$, which is estimated by profile maximum likelihood, but including the nugget variance $\tau^2$, the four models have 3, 4, 6, and 12 unknown parameters. Each model has a multiplicative variance parameter $\sigma^2$. In the Nelder-Mead algorithm, we profile out $\sigma^2$, whereas In Fisher scoring, we do not. We found that profiling $\sigma^2$ does not substantially influence convergence speed in Fisher scoring. All positive parameters are mapped to the real line by a log transform. Each model was fit to each dataset on an independent R process running on a single thread of an 8-core (16 thread) Intel Xeon W-2145 (3.7GHz, 4.5GHz Turbo) processor with 16GB RAM. Fifteen datasets were sent to each of 14 threads, yielding 210 simulated datasets.

\begin{figure}
\centering
\includegraphics[width=1.0\textwidth]{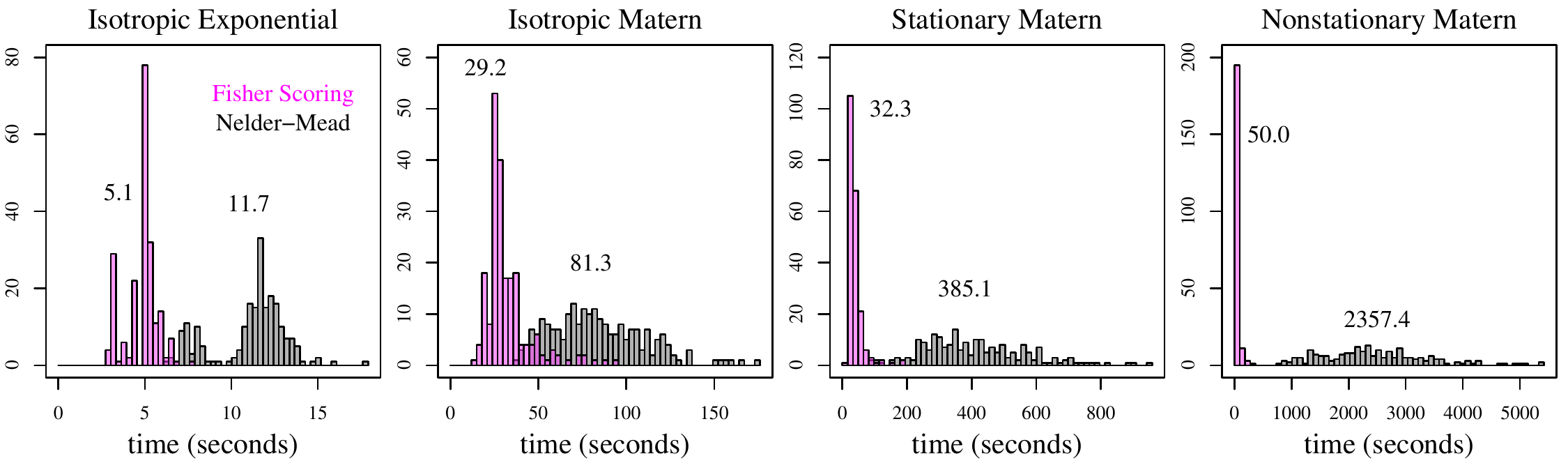}
\vspace{-20pt}
\caption{\label{timing_results} Results of optimization timing study. Each plot shows histograms of time (in seconds) until convergence for one of the four covariance functions over 210 replicates. The plotted numbers indicate the median time until convergence.}
\end{figure}

Hisotograms of the timing results are given in Figure \ref{timing_results}. Considering the median times, Fisher scoring is 2-3 times faster for the isotropic models (3 and 4 parameters), more than 10 times faster for the stationary Mat\'ern model (6 parameters), and 47 times faster for the nonstationary model (12 parameters). There is also no noticeable loss in accuracy, which we can evaluate by comparing the maximum loglikelihoods returned by Fisher scoring to the loglikelihoods returned by Nelder-Mead. For the isotropic models, the two loglikelihoods never differed by more than 0.001 units. In the stationary model, the Fisher scoring loglikelihood was more than 0.01 larger than the Nelder-Mead loglikelihood in 11 of the 210 datasets, whereas the Nelder-Mead loglikelihood was never more than 0.01 larger than the Fisher scoring loglikelihood. For the nonstationary model, the Fisher scoring loglikelihoods were more than 0.01 larger than the Nelder-Mead loglikelihoods in 199 of the 210 datasets, whereas the Nelder-Mead loglikelihoods were more than 0.01 larger in just 3 of the 210 datasets. Not only does Nelder-Mead take nearly 50 times longer than Fisher scoring in the nonstationary model, it usually does not find the actual maximum likelihood estimates.

\subsection{Identifiability}

Surprisingly, the maximum likelihood estimate of the nugget variance can be a negative number. This is not an error of the optimization algorithm (a triumph rather); in these cases, the algorithm finds a negative nugget with a higher likelihood than any nonnegative nugget can produce. Negative nugget estimates occur most frequently when the data are evenly spaced and when the maximum likelihood estimate of the smoothness parameter is small ($\nu < 0.25$). In this scenario, the covariance function has a narrow peak at distances smaller than the minimum spacing between locations. This happened frequently enough in our testing of the Fisher scoring algorithm that we found it necessary to impose a penalty on very small values of the nugget and smoothness parameters, since it is not sensible to return negative nugget estimates. The penalties are
\begin{align*}
\mbox{pen}(\tau^2) = - 0.01 \log( 1 + 0.01/\tau^2 ), \quad
\mbox{pen}(\nu) = - 0.01 \log( 1 + 0.2/\nu ).
\end{align*}
The likelihood function also has difficulty jointly identifying variance and range parameters when the range parameter estimate is much larger than the maximum distance between points in the dataset. This is a theoretically well-studied problem \citep{zhang2004inconsistent} that no optimization routine can overcome. We have found that penalizing large variance parameters helps improve convergence of Fisher scoring without sacrificing accuracy. We used the penalty
\begin{align*}
\mbox{pen}(\sigma^2) = \log( 1 + e^{\sigma^2/\widetilde{\sigma}^2 - 6} ),
\end{align*}
where $\widetilde{\sigma}^2$ is the estimate of the residual variance parameter in a least squares fit of the response to the constant covariate. This imposes essentially no penalty on the parameter unless it is several times larger than the least squares estimate, after which the penalty increases roughly linearly in $\sigma^2$. These two identifiability problems can be handled more elegantly in a Bayesian framework, but we do not pursue that here because identifiability is not the focus of this paper.

\section{Case Study: Argo Ocean Temperature Data}

\begin{figure}
\centering
\includegraphics[width=\textwidth]{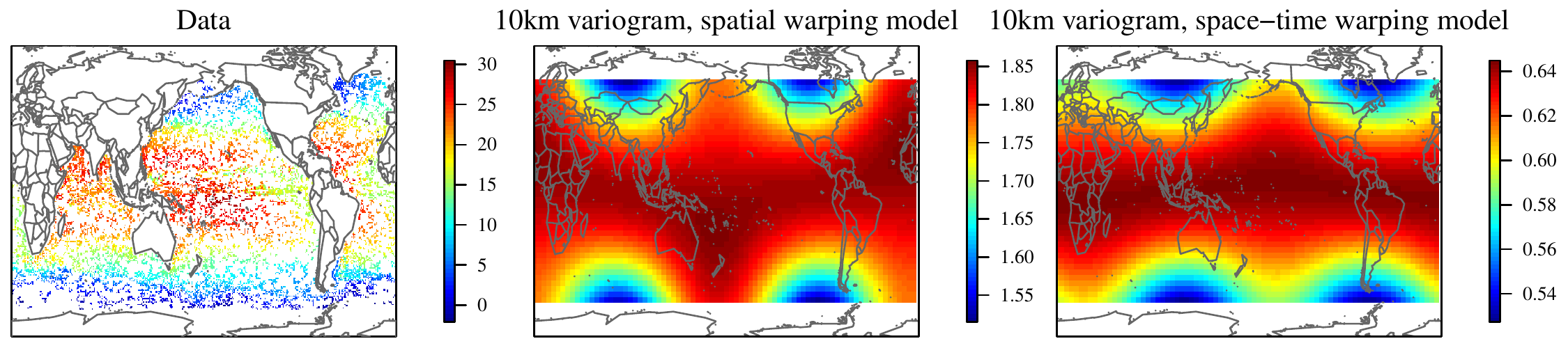}
\vspace{-20pt}
\caption{\label{argo_figure} Plot of Argo data and variograms evaluated at 10km
for the spatial-only model, and the spatial-temporal model.}
\end{figure}

Argo is a global program that deploys floating ocean temperature sensors \citep{argo}. Each
Argo float operates on a 10 day cycle, during which it descends to a 2000m depth
and returns to the surface, collecting temperature and salinity measurements
along the depth profile. The floats drift freely in the horizontal direction with
ocean currents.  As of May 2019, 3{,}799 floats covered the
globe. We analyze a subset of the observations collected at 100
dbar (approximately 100m depth) between January 1 and March 31, 2016. Preprocessed data were
provided by Mikael Kuusela and are described in more detail in
\cite{kuusela2018locally}. In total, there are
32{,}492 measurements over the three month period. The data are plotted in Figure \ref{argo_figure}.

We model the data from day $t$ and location $s$ on the sphere $\mathbb{S} \subset \R^3$ as
\begin{align*}
Y(s,t) = \beta_0 + \beta_1 L(s) + \beta_2 L^2(s) + Z(s + \Phi(s),t) + \varepsilon(s,t),
\end{align*}
where $L(s)$ is the latitude of location $s$, $Z(s,t)$ is a Gaussian process
with covariance function $K_\theta$, $\Phi: \R^3 \to \R^3$ is a spatial warping
function, and $\varepsilon(s,t)$ are i.i.d.\ mean zero
normals with variance $\tau^2$. We consider both spatial and spatial-temporal
models for $K_\theta$:
\begin{align*}
K_\theta((s_1,t_1),(s_2,t_2)) &= \frac{\sigma^2}{2^{\nu-1} \Gamma(\nu)}
\left( d_{\alpha}(s_1,s_2) \right)^\nu \mathcal{K}_\nu\left( d_{\alpha}(s_1,s_2) \right), \\
K_\theta((s_1,t_1),(s_2,t_2)) &= \frac{\sigma^2}{2^{\nu-1} \Gamma(\nu)}
\left( d_{\alpha}((s_1,t_1),(s_2,t_2)) \right)^\nu \mathcal{K}_\nu\left( d_{\alpha}((s_1,t_1),(s_2,t_2)) \right).
\end{align*}
The function $d_\alpha$ is Euclidean distance scaled by either a spatial
range parameter or spatial and temporal range parameters
\begin{align*}
d_{\alpha}(s_1,s_2) &= \frac{\| s_1 - s_2 \|}{\alpha}, \quad d_{\alpha}((s_1,t_1),(s_2,t_2)) = \left( \frac{\| s_1- s_2 \|^2}{\alpha_1^2} + \frac{|t_1-t_2|^2}{\alpha_2^2} \right)^{1/2}.
\end{align*}
The warping function $\Phi$ is assumed to be a linear combination of the
gradients of the five spherical harmonic functions of degree 2,
where the gradient is with respect to the three Euclidean coordinates. We use degree 2
because the degree 0 function is constant, and the degree 1 spherical harmonics
have constant partial derivatives (as a function of $s$), and so degree 1
warpings simply translate all points by the same vector and do not affect
the covariances. We also consider the special case of $\Phi(s) = 0$ for all $s$,
which corresponds to isotropic models in space and time. The spatial warping model has 9 parameters, while the space-time warping model has 10. The isotropic models have 4 and 5 parameters.

We fit each model using both Fisher scoring and Nelder-Mead,
with the results given in Table \ref{argo_table}.
Fisher scoring is able to fit the space-time warping model
in 12.13 minutes, whereas Nelder-Mead ran for 190 minutes and
returned a loglikelihood value 2.038 units lower. In the spatial-only warping
model, Fisher Scoring finished in 6.56 minutes, whereas Nelder-Mead returned
a loglikelihood value 0.01 lower after 80 minutes. The two methods produced
nearly the same loglikelihoods on the isotropic models, with Fisher scoring
running more than twice as fast. The results closely mirror the
numerical study, where Fisher scoring had its largest
improvements in both speed and accuracy when fitting models with the many parameters.
Finally, in Figure \ref{argo_figure}, we plot $\Var(Y(s,t)-Y(s+h,t))$ as a function of $s$,
with $\|h\| = 10$km. The images show that the warping model produces an
anisotropic variogram, with larger increment variances near the equator.

\begin{table}
\centering
\begin{tabular}{crrrr}
 & \multicolumn{2}{c}{loglikelihood} & \multicolumn{2}{c}{time (minutes)} \\
 Model & Fisher Scoring & Nelder-Mead & Fisher Scoring & Nelder-Mead \\
\hline
Isotropic Spatial &$-6167.675$ & $-6167.676$ & 2.76 & 7.06 \\
Warping Spatial &$-5812.902$ & $-5812.912$ & 6.56 & 79.96 \\
Isotropic Space-Time &$-237.038$ & $-237.039$ & 3.62 & 10.61 \\
Warping Space-Time &$0.000$ & $-2.038$ & 12.13 & 190.31\\
\end{tabular}
\caption{\label{argo_table} Optimization results for four models fit to Argo
float data. Reported loglikelihoods are differences from largest loglikelihood}
\end{table}

\section{Discussion}

We believe that practitioners will benefit from the
availability of high quality algorithms for fitting nonstationary Gaussian
process models to large spatial and spatial-temporal datasets. The methods are applicable
to any covariance function that is differentiable with respect to its parameters. This is important because it separates the tasks
of constructing models and developing methods for fitting the models, freeing us to select the most
appropriate covariance function for the data rather than the most appropriate model
for which a specialized method exists.
The Fisher scoring algorithm, as well as anisotropic, nonstationary variance,
and warping covariance functions, will be implemented in version 0.2.0 of the GpGp
R package \citep{GpGp}.

\begin{center}
\Large{\bf Acknowledgements}
\end{center}

This work was supported by the National Science Foundation under grant No.\ 1613219 and the National Institutes of Health under grant No.\ R01ES027892.

\bibliography{refs}{}

\begin{thebibliography}{}

\bibitem[Finley et~al., 2017]{spNNGP}
Finley, A., Datta, A., Banerjee, S., and Mckim, A. (2017).
\newblock {\em spNNGP: Spatial Regression Models for Large Datasets using
  Nearest Neighbor Gaussian Processes}.
\newblock R package version 0.1.1.

\bibitem[Geoga et~al., 2018]{geoga2018scalable}
Geoga, C.~J., Anitescu, M., and Stein, M.~L. (2018).
\newblock Scalable {G}aussian process computations using hierarchical matrices.
\newblock {\em arXiv preprint arXiv:1808.03215}.

\bibitem[Guinness, 2018]{guinness2018permutation}
Guinness, J. (2018).
\newblock Permutation and grouping methods for sharpening {G}aussian process
  approximations.
\newblock {\em Technometrics}, 60(4):415--429.

\bibitem[Guinness and Katzfuss, 2018]{GpGp}
Guinness, J. and Katzfuss, M. (2018).
\newblock {\em GpGp: Fast Gaussian Process Computation Using Vecchia's
  Approximation}.
\newblock R package version 0.1.0.

\bibitem[{International Argo Program}, 2019]{argo}
{International Argo Program} (2019).
\newblock \url{http://www.argo.ucsd.edu/}.
\newblock Online: accessed 2019-05-19.

\bibitem[Katzfuss and Guinness, 2017]{katzfuss2017general}
Katzfuss, M. and Guinness, J. (2017).
\newblock A general framework for {V}ecchia approximations of {G}aussian
  processes.
\newblock {\em arXiv preprint arXiv:1708.06302}.

\bibitem[Kuusela and Stein, 2018]{kuusela2018locally}
Kuusela, M. and Stein, M.~L. (2018).
\newblock Locally stationary spatio-temporal interpolation of {A}rgo profiling
  float data.
\newblock {\em Proceedings of the Royal Society A}, 474(2220):20180400.

\bibitem[Nelder and Mead, 1965]{nelder1965simplex}
Nelder, J.~A. and Mead, R. (1965).
\newblock A simplex method for function minimization.
\newblock {\em The computer journal}, 7(4):308--313.

\bibitem[Stein et~al., 2004]{stein2004approximating}
Stein, M.~L., Chi, Z., and Welty, L.~J. (2004).
\newblock Approximating likelihoods for large spatial data sets.
\newblock {\em Journal of the Royal Statistical Society: Series B (Statistical
  Methodology)}, 66(2):275--296.

\bibitem[Vecchia, 1988]{vecchia1988estimation}
Vecchia, A.~V. (1988).
\newblock Estimation and model identification for continuous spatial processes.
\newblock {\em Journal of the Royal Statistical Society: Series B
  (Methodological)}, 50(2):297--312.

\bibitem[Zhang, 2004]{zhang2004inconsistent}
Zhang, H. (2004).
\newblock Inconsistent estimation and asymptotically equal interpolations in
  model-based geostatistics.
\newblock {\em Journal of the American Statistical Association},
  99(465):250--261.

\end{thebibliography}
\bibliographystyle{apalike}

\appendix

\end{document}